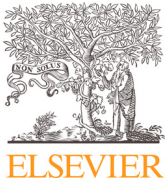

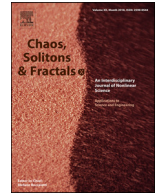

# Multifractal fluctuations of the precipitation in Spain (1960–2019)

Javier Gómez-Gómez\*, Rafael Carmona-Cabezas, Elena Sánchez-López,
Eduardo Gutiérrez de Ravé, Francisco José Jiménez-Hornero

*GEPENA Research Group, University of Cordoba, Gregor Mendel Building (3rd floor), Campus Rabanales, Cordoba 14071, Spain*

ABSTRACT

In this work, an analysis of multifractal parameters of daily precipitation series over the Iberian Peninsula was performed in two 30-year periods to explore whether these properties follow any pattern. Fluctuations of precipitation series show three different scaling regions. Only two distinct regimes for small and large timescales can be confirmed, while intermediate scales are part of a transition region. It is also observed a certain degree of multifractality, which is higher for small timescales. At these scales, there is a high persistence which follows the spatial gradient of the annual precipitation. Moreover, multifractal parameters of the precipitation are modified according to complex spatial and temporal patterns. Only persistence uniformly decreases in the last period. Other relevant findings are the changes in the asymmetry of multifractal spectra in the eastern belt at larger timescales, which might be related to the change in the behavior of the Mediterranean cyclones.

© 2022 The Authors. Published by Elsevier Ltd.
This is an open access article under the CC BY-NC-ND license
(http://creativecommons.org/licenses/by-nc-nd/4.0/)

## 1. Introduction

Climate change might impact natural and human systems in different ways. Some include changes in quantity and quality of water resources, negative impacts on agriculture or an increment in frequency of droughts, floods or wildfires, among others [1,2]. A closely relation between some of these effects and precipitation patterns, makes quite relevant the study of this climatic variable for a better understanding of the hydrological systems [3].

Precipitation is a climatic variable of great relevance for the Iberian Peninsula. This region is located between Africa and Central Europe, more precisely, at the south of Europe and in the western Mediterranean basin. Unlike other Mediterranean regions, this one is characterized by having the Atlantic oceanic influence, what produces a climate with irregularity in water regime [4], which makes this geographic location specially interesting to study rainfall. According to de Luis et al. [5], seasonal precipitation regimes in the Iberian Peninsula can be explained in a simple way, as the sum of three factors that contributes to the high spatial variability: an Iberian inland component, and the Atlantic and Mediterranean oceanic components which becomes more notable at the west and east, respectively. Thus, water resources scarcity is notable in this region, as it is shown by some models that predict a drier precipitation regime, particularly, due to a prolonged dry season [6]. Ac-tually, several authors showed that seasonal precipitation regimes are modifying in this region [5,7–9].

Climate variability studies based on empirical meteorological data are of great importance. Much of them are based on statistical analysis [10–13]. Nevertheless, an growing interest in multifractal analysis is taking place in the last decades [14–16]. Multifractal analysis is useful to study the complexity and non-linearity of time series which cannot be addressed with other linear methods. These techniques are based on the fractal theory [17,18]. A fractal is a geometric object characterized by its self-similarity or scale independency when they are split into smaller parts. Many studies have demonstrated that fluctuations of several environmental variables and, particularly, precipitation, have multifractal nature, and contain a range of scaling exponents which characterize the temporal structure of the time series. Thus, the underlying process can be described by the multifractal parameters. Some of the most important ones studied here are the Hurst exponent, the Hölder exponent with maximum spectrum, the width and the asymmetry of the multifractal spectrum [14,19,20].

In recent years, a technique is being widely used to get information about the multifractal scaling properties of temporal fluctuations in signals. This is the multifractal detrended fluctuation analysis (MF-DFA), which was developed by Kantelhardt et al. [21]. This technique is a useful method to deal with non-stationary time series. Its applications extend to several climatic variables, as exposed in different studies [3,16,20,22].

Due to the high variability in the spatial and temporal distribution of rainfall in the Iberian Peninsula, it is appropriate to analyze the complexity and non-linearity of precipitation series in dif-ferent time periods. Thus, it will be possible to explore whether these properties follow geographical or temporal patterns. To this purpose, two independent 30-year periods are analyzed with MF-DFA and compared. Because of the data availability and their quality, the two 30-year periods are limited to 1960 – 1989 and 1990 – 2019. Hence, the analysis of the spatial and temporal variability is performed through the study of the changes in multifractal parameters of daily precipitation series between these periods. Furthermore, a preliminary analysis of linear trends of annual precipitation series is carried out to improve the information extracted about the climate variability in this region. This work is structured as follows. Section 2 includes the description of data, gauge stations and methods used; results are collected and discussed in Section 3 and, lastly, Section 4 includes the main conclusions drawn in this study.

\* Corresponding author.
  *E-mail address:* f12gogoj@uco.es (J. Gómez-Gómez).

## 2. Materials and methods

### 2.1. Data

This study is based on daily precipitation data series from 29 meteorological stations distributed over the Spanish region of the Iberian Peninsula during the period 1960 - 2019 (see Fig. 1). This leads to a total amount of 21,915 records, split up into two sub-periods (1960 - 1989 and 1990 - 2019) with $N_1 = 10958$ and $N_2 = 10957$, respectively.

Raw data records were obtained via the Spanish Meteorological Agency ("Agencia Estatal de Meteorología") from the AEMET Open-Data website at http://www.aemet.es/es/datos_abiertos/AEMET_OpenData. This network contains 261 stations across the considered region. To get reliable results, a previous identification of missing data was carried out and we discarded time series which contained more than 0.01% missing values and more than 10 consecutive ones for both periods. After this procedure, the remaining 29 series were collected. Some basic information about these stations can be seen in the Appendix A, in Table A.1. The annual precipitation was also computed from daily rainfall and an analysis of inter-annual trends was performed through simple least-squares fits.

The set of gauge stations covers the main two climate variants of the Iberian Peninsula: the Atlantic climate type and the Mediterranean semiarid subtype [23]. The precipitation climatology in the Spanish region of the Iberian Peninsula is characterized by strong gradients with abundant annual precipitation to the north and northwest ($P > 1000$ mm/year) and lower values towards the southeast ($P < 400$ mm/year) [10]. In this last region, certain areas can even have less than 200 mm/year [24]. However, a higher complexity in the spatial and temporal seasonal variability is present in the study area, because of the two water masses influences and orography [5]. Fig. 1(b) shows the orography of the region.

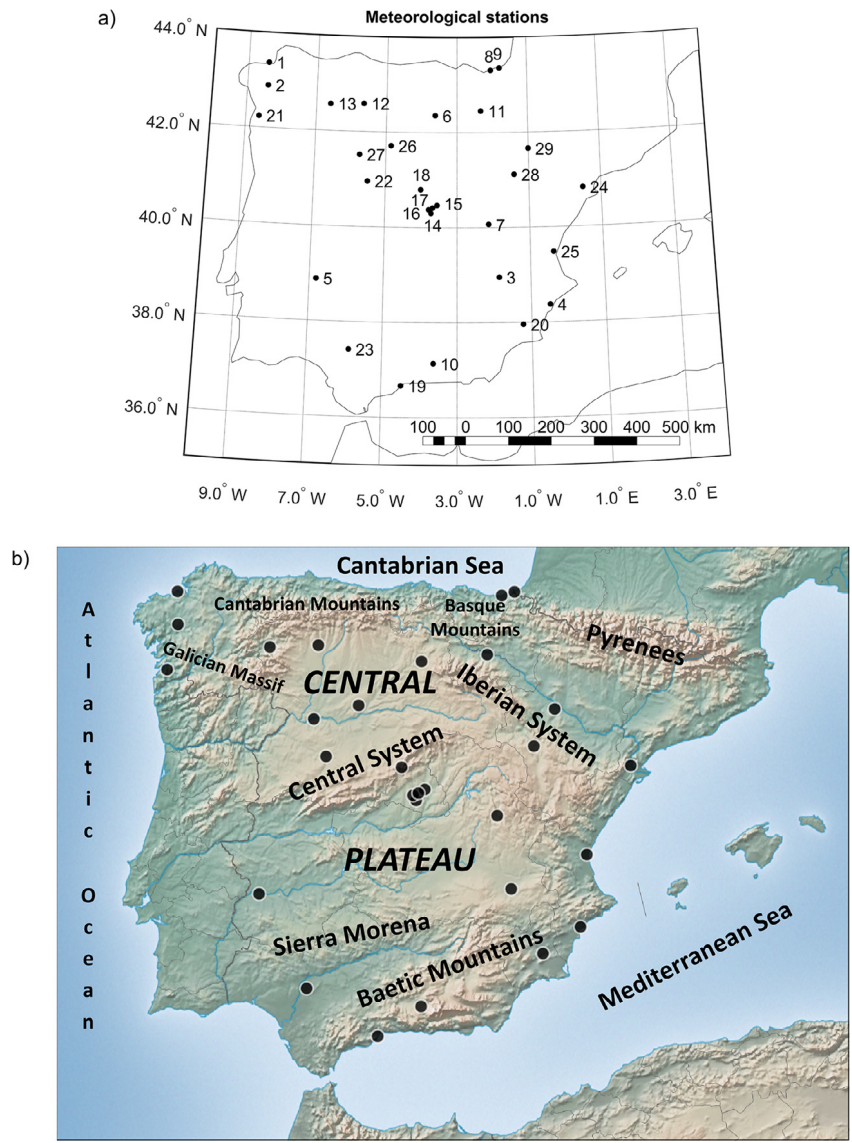

**Fig. 1.** (a) Meteorological stations. (b) Orographic map of the Iberian Peninsula.

## 2.2. Multifractal detrended fluctuation analysis

In order to apply the MF-DFA algorithm, it is common to contemplate the deseasonalized time series when analyzing climatological variables [16,25] and, in particular, hydrological variables [26–28]. In our study, the precipitation (normalized) anomalies were computed similarly to the work of Xavier et al. [29], to avoid that seasonal periodicity affected power law behavior. As an example, the whole time series with daily resolution of station No. 15 and its corresponding deseasonalized version are shown in Fig. 2.

MF-DFA algorithm was developed by Kantelhardt et al. [21]. This method computes the "profile" (the integrated series after subtracting the mean) and obtains its fluctuation function dividing this profile into multiple segments. More precisely, the profile is divided into $N_s \equiv int(N/s)$ nonoverlapping segments of equal length $s$ and the same procedure is repeated from the end of the series to the beginning. As a result, $2N_s$ segments are obtained altogether for each time scale $s$ (see [21] for more details). Next, the local trend, $y_v(i)$, must be determined for each segment $v$ by the least-squares fit of the values [30,31]. In this document, the simple linear fit has been used. The local trend is subtracted from the

profile as follows:

$$F^2(v, s) \equiv \frac{1}{s} \sum_{i=1}^{s} \{Y[(v-1)s + i] - y_v(i)\}^2 \qquad (1)$$

for each segment $v$, $v = 1, \ldots, N_s$ and

$$F^2(v, s) \equiv \frac{1}{s} \sum_{i=1}^{s} \{Y[N - (v - N_s)s + i] - y_v(i)\}^2 \qquad (2)$$

for each segment $v$, $v = N_s + 1, \ldots, 2N_s$.

Finally, to compute the $q$th order fluctuation function, the average over all segments is computed:

$$F_q(s) \equiv \left\{ \frac{1}{2N_s} \sum_{v=1}^{2N_s} \left[ F^2(v, s) \right]^{q/2} \right\}^{1/q} \qquad (3)$$

Since the averaging procedure in Eq. (3) cannot be used for $q = 0$, a logarithmic averaging procedure must be applied instead [21].

If the analyzed series are long-range power law correlated, $F_q(s)$ increases for large $s$ as a power-law:

$$F_q(s) \sim s^{h(q)} \qquad (4)$$

Consequently, the scaling exponent $h(q)$ can be obtained by means of the computation of slopes in the log-log plots of $F_q(s)$ vs $s$ for each $q$.

Negative $q$ values are related to the intervals with small fluctuations whereas positive ones describe the scaling behavior of large fluctuations [32]. In general, $h(q)$ can depend on $q$, meaning that small and large fluctuations significantly scale in different ways. Only if $h(q)$ is independent of $q$, the series has monofractal nature.

For stationary signals, $h(2)$ is the Hurst exponent $H$ whereas, for non-stationary signals, it is retrieved from $H = h(2) - 1$ [33]. For this reason, $h(q)$ is called as generalized Hurst exponent [21,26,34]. The standard Hurst exponent, $H$, provides information about the long-range correlations of the signals [34].

## 2.3. Relation to standard multifractal analysis

Kantelhardt et al. obtained an analytical expression [21] that connected the MF-DFA to the standard box counting formalism [17,18]. As they demonstrated, the scaling exponent $h(q)$ defined in Eq. (4) is related to the scaling exponent $\tau(q)$, which is determined by the partition function of the multifractal formalism. Therefore, the multifractal spectrum or singularity spectrum, $f(\alpha)$, can be computed via the Legendre transform as [35]:

$$\alpha = \frac{d\tau(q)}{dq} \text{ and } f(\alpha) = q\alpha - \tau(q) \qquad (5)$$

where $\alpha$ is the singularity strength or Hölder exponent [36].

The shape of $f(\alpha)$ is often a concave-down parabola with a maximum value which is the most dominant scaling behavior [20]. The singularity strength at which $f(\alpha)$ reaches its maximum value is usually denoted by $\alpha_0$. Large values of this parameter indicates that the underlying process has "fine-structure" and is more complex [15,37]. Another important quantity from this spectrum is its width, $w$, which provides information about the degree of multifractality of the signal [16]. A monofractal time series will have a spectral width close to zero.

There are different kinds of parameters in literature to measure the asymmetry of multifractal spectra [15,20,37–39]. In this manuscript, it is used one based on the work of Shimizu et al. [37], which takes the multifractal spectrum and compute the second order polynomial fit of the shifted curve:

$$f(\alpha) = A(\alpha - \alpha_0)^2 + B(\alpha - \alpha_0) + C \qquad (6)$$

The $B$ coefficient is usually known as the asymmetry parameter. When this parameter is equal to zero, the spectrum itself is symmetrical. On the other hand, if $B > 0$ the curve is left-skewed and it is right-skewed when $B < 0$ [19,37]. A right-skewed spectrum indicates that small fluctuations ($q < 0$) are favored and the time series is less singular and has "fine-structure" [37,40]. A left-skewed shape is associated to time series which are more singular and are characterized by a "loss of fine-structure".

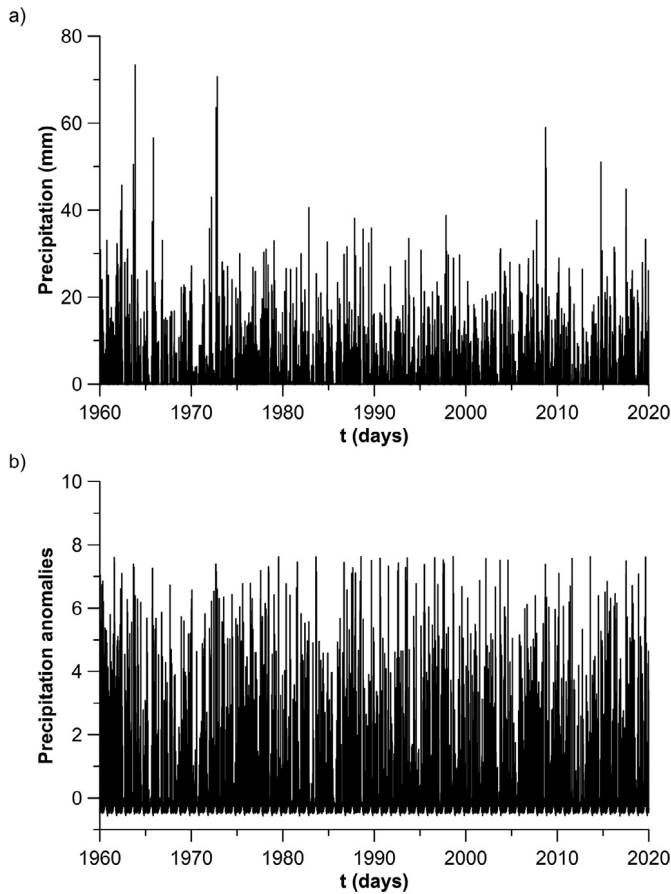

**Fig. 2.** Complete time series and its corresponding deseasonalized version (precipitation anomalies) of station No. 15.

## 3. Results and discussion

### 3.1. Descriptive statistics and trend analysis of annual precipitation

Annual precipitation series were computed from daily records and linear fits, based on least-squares regressions, were applied to these series as a preliminary study of inter-annual variability. The presence of significant linear trends was analyzed for each location. Results are given in Table 1. To this aim, t-tests were performed.

A total amount of 14 stations (48% of sample) shows statistically significant negative trends in the whole period 1960 - 2019. An example of four stations is depicted in Fig. 3, where trends are statistically significant in the considered period. In these stations, trends of $-10.90$, $-1.71$, $-4.10$ and $-2.59$ mm/year, respectively, were found. The most pronounced decreasing trend corresponds to the station No. 2 (Fig. 3(a)), located at the northwestern coastal subregion of the Iberian Peninsula, which is characterized by abundant precipitation. The second highest decreasing trend belongs to the station No. 21, which is also located at this region. The rest of locations with significant trends belong to the central and southern mainland subregions, except for No. 24 (Fig. 3(d)), which is at the northeastern Mediterranean coastland. Furthermore, all these 14 significant trends are consistent with a decrease in mean and standard deviation of annual precipitation for the same stations between both subperiods 1960 - 1989 and 1990 - 2019 (see Table 1).

### 3.2. Hurst exponents for different regions and for periods 1960 - 1989 and 1990 - 2019

MF-DFA method was applied to deseasonalized daily precipitation series at 29 locations for each subperiod. The $q$ dependent fluctuation function was obtained from these signals for $q \in [-5, 5]$ with step 0.5 and over a wide range of time scale $s$, more precisely, from $s = 10$ to 1094 days ($\approx N/10$) and step of 2. Consequently, log-log plots of $F_q(s)$ vs $s$ were obtained for each case. Plots for stations No. 15, 20 and 2 in both subperiods are depicted in Fig. 4. It can be observed that the fluctuation function has three clearly differentiated scaling regions, indicated by (i) - (iii). All stations exhibit these three distinct regimes. However, the amplitudes of these regions vary, and might be shifted depending on the location (see Fig. 4). Moreover, the same location can exhibit different limits for the three regions ((i) – (iii)) in both subperiods of 30 years. Average values of crossovers that separate the three scaling regimes are the following. The average first crossover that separate regions (i) and (ii) is 45 and 41 days for the first and second subperiods, respectively. The second average crossovers are 151 and 134 days.

The power spectra of time series only show two different scaling regions which can be clearly distinguished, as it can be seen for the above-mentioned stations in Fig. 4(g), (h). Most stations display statistically non-significant slopes for low frequencies, while all of them show marked slopes for high frequencies. This means that for a wide range of high frequencies, the precipitation series show a power-law decay of the spectrum, $P(f) \propto f^{-\beta}$, what indicates a strong persistence at small scales [41]. The linear fits in this last interval of frequencies agree the region (i) found for $s$. As a consequence, both results confirm that fluctuations of precipitation series present mainly two different scaling regimes for large and small timescales, while the region (ii) seems to be a transition region for small fluctuations ($q < 0$).

**Table 1**
Mean, standard deviation (SD) and daily maximum (Max.) of precipitation for each subperiod, together with the slope and Pearson correlation coefficient of linear fits obtained by least-squares regression of annual precipitation series. Slopes highlighted in bold are statistically significant at 95% confidence level.

| No. | 1960–1989 | | | 1990–2019 | | | Annual trends | |
| --- | Mean | SD | Max. | Mean | SD | Max. | Slope (mm/year) | R |
| 1 | 1014.1 | 169.5 | 66.6 | 1017.0 | 175.7 | 132.7 | $-0.06 \pm 1.3$ | 0.01 |
| 2 | 1975.4 | 400.8 | 218.0 | 1671.3 | 366.4 | 118.6 | $\mathbf{-10.90 \pm 2.7}$ | 0.46 |
| 3 | 370.2 | 98.0 | 78.6 | 351.5 | 101.0 | 146.6 | $-0.27 \pm 0.7$ | 0.05 |
| 4 | 359.6 | 116.9 | 220.2 | 286.3 | 114.7 | 270.2 | $-1.22 \pm 0.9$ | 0.18 |
| 5 | 507.8 | 134.8 | 70.5 | 424.8 | 130.5 | 119.1 | $\mathbf{-2.54 \pm 1.0}$ | 0.32 |
| 6 | 584.7 | 111.4 | 51.6 | 545.1 | 96.9 | 52.4 | $\mathbf{-1.71 \pm 0.8}$ | 0.28 |
| 7 | 577.9 | 151.2 | 69.6 | 490.9 | 115.0 | 98.2 | $\mathbf{-2.84 \pm 1.0}$ | 0.35 |
| 8 | 1590.5 | 242.1 | 149.8 | 1560.2 | 221.1 | 167.7 | $-0.20 \pm 1.7$ | 0.02 |
| 9 | 1747.0 | 268.5 | 130.4 | 1686.3 | 257.3 | 185.2 | $-2.29 \pm 2.0$ | 0.15 |
| 10 | 390.6 | 89.0 | 53.6 | 356.3 | 117.0 | 69.3 | $-0.93 \pm 0.8$ | 0.16 |
| 11 | 391.1 | 78.9 | 64.6 | 429.3 | 87.0 | 82.6 | $0.71 \pm 0.6$ | 0.15 |
| 12 | 568.5 | 116.3 | 98.5 | 495.5 | 109.3 | 56.5 | $\mathbf{-1.92 \pm 0.8}$ | 0.28 |
| 13 | 668.0 | 146.6 | 58.3 | 638.1 | 140.9 | 63.6 | $-0.50 \pm 1.1$ | 0.06 |
| 14 | 429.4 | 100.7 | 62.0 | 352.3 | 80.5 | 64.6 | $\mathbf{-2.32 \pm 0.7}$ | 0.41 |
| 15 | 423.2 | 133.8 | 73.4 | 367.4 | 95.2 | 58.9 | $\mathbf{-1.89 \pm 0.9}$ | 0.28 |
| 16 | 472.4 | 111.6 | 66.8 | 411.7 | 93.4 | 79.4 | $\mathbf{-2.04 \pm 0.8}$ | 0.34 |
| 17 | 464.3 | 121.3 | 87.0 | 411.1 | 95.8 | 50.2 | $\mathbf{-1.65 \pm 0.8}$ | 0.26 |
| 18 | 1438.5 | 323.0 | 116.0 | 1254.6 | 296.5 | 150.0 | $-4.61 \pm 2.3$ | 0.25 |
| 19 | 594.2 | 237.3 | 151.0 | 502.2 | 237.7 | 132.7 | $-3.14 \pm 1.8$ | 0.23 |
| 20 | 296.5 | 117.5 | 99.8 | 287.4 | 92.7 | 179.7 | $0.34 \pm 0.8$ | 0.06 |
| 21 | 2001.6 | 357.3 | 175.0 | 1690.4 | 347.4 | 171.9 | $\mathbf{-9.89 \pm 2.6}$ | 0.45 |
| 22 | 395.8 | 88.2 | 48.7 | 354.6 | 76.9 | 50.3 | $\mathbf{-1.34 \pm 0.6}$ | 0.28 |
| 23 | 623.3 | 202.7 | 101.0 | 503.3 | 171.5 | 109.3 | $\mathbf{-4.10 \pm 1.4}$ | 0.37 |
| 24 | 568.3 | 192.8 | 176.5 | 499.6 | 111.4 | 140.8 | $\mathbf{-2.59 \pm 1.2}$ | 0.28 |
| 25 | 465.1 | 187.2 | 148.4 | 454.4 | 154.0 | 178.2 | $0.22 \pm 1.3$ | 0.02 |
| 26 | 486.9 | 109.6 | 80.0 | 422.6 | 108.8 | 60.5 | $\mathbf{-1.88 \pm 0.8}$ | 0.29 |
| 27 | 397.6 | 121.9 | 66.1 | 388.0 | 89.8 | 51.4 | $-0.67 \pm 0.8$ | 0.11 |
| 28 | 434.3 | 97.8 | 76.6 | 392.9 | 95.7 | 68.8 | $-0.94 \pm 0.7$ | 0.17 |
| 29 | 315.3 | 77.3 | 67.3 | 326.9 | 88.1 | 70.8 | $0.56 \pm 0.6$ | 0.12 |

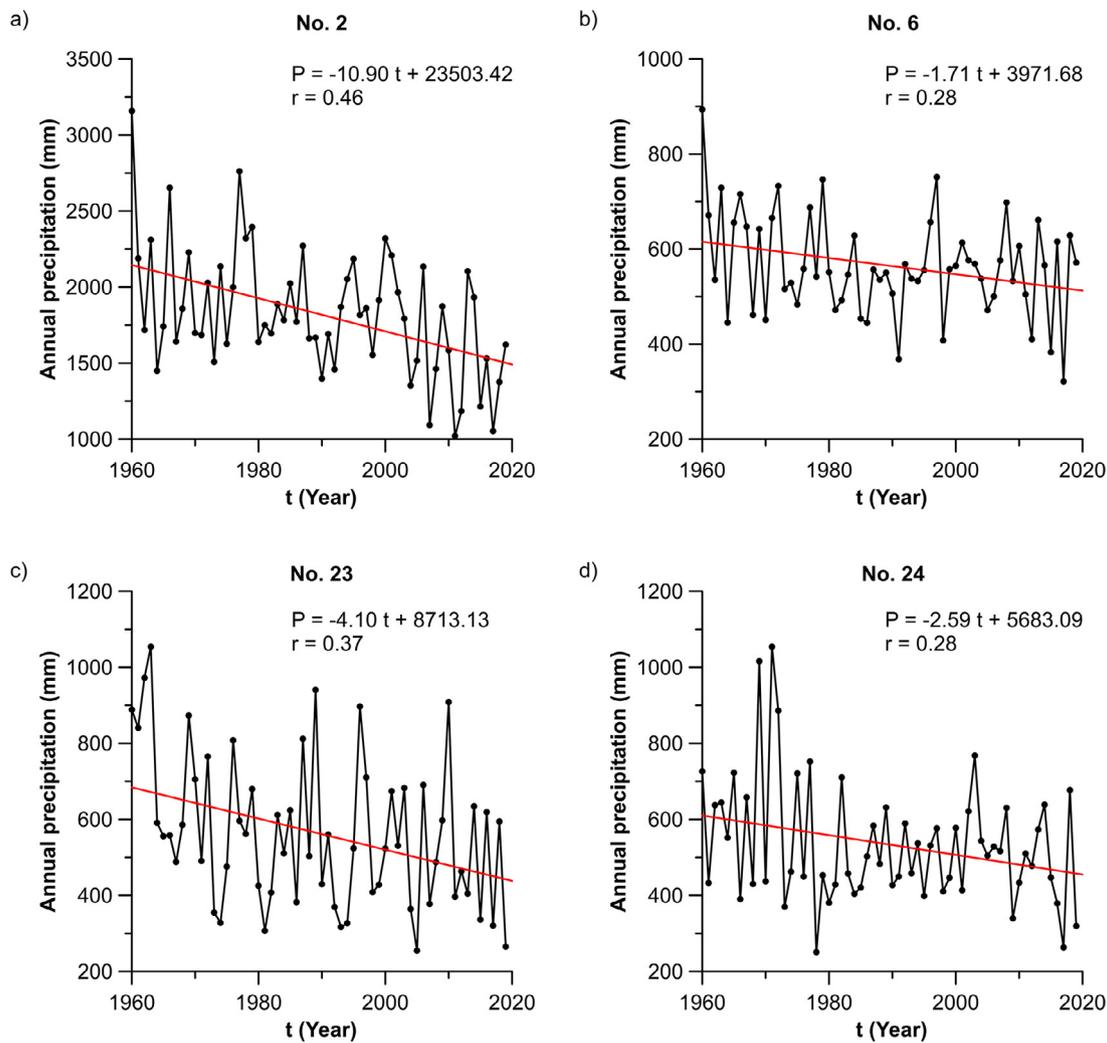

**Fig. 3.** Linear fits of annual precipitation in the period 1960–2019 of four meteorological stations.

The best linear fits of $F_q(s)$ in station No. 15 for regions (i) and (iii) are depicted in Fig. 5(a). Generalized Hurst exponents, $h(q)$, obtained from these linear fits are shown against $q$ moments in Fig. 5(b). A non-monotonic function of $q$ can be seen in the region (i) due to the abrupt change of the tendency for negatives $q$ values (see Fig. 5(a), (b)). This phenomenon might be related to the limited resolution of the daily time series which might be causing spurious results when the small fluctuations are magnified by the method in these smaller scales. Similar outcomes have been obtained for the rest of stations.

A clear evidence of multifractality is the $q$ dependence of $h$. In fact, a measurement of the degree of multifractality is the range of $h(q)$ [32]. There is a wide range of $h$ for small timescales and a small but significant range of values for large timescales, as it is shown in the Fig. 5(b). Therefore, these signals are multifractals and show a greater degree of multifractality at small timescales.

Regarding the well-known Hurst exponent, $H$, some interesting results have been obtained. All values are lower than 1, what indicates that time series are stationary [33]. The greatest values of $H$ correspond to the smallest timescales (region (i)), as it is also confirmed by the marked slopes found in the power spectra. For this regime, $H_i \in [0.58, 0.79]$, while the region (iii) shows values of $H_{iii} \in [0.38, 0.65]$. These last weak long-term correlations depend on the location and only station No. 16 for the last period has a Hurst exponent similar to a white noise process considering results obtained from MF-DFA. Nevertheless, power spectra at these timescales only exhibit statistically significant slopes for stations No. 3, 8, 13 and 19 for the first period and No. 1, 2, 11, 18 and 21 for the second one. Thus, all precipitation series show a regime with strong long-term persistence for small timescales and other regime where they resemble white noise for large timescales. Due to this discrepancy in the results, the analysis of the possible changes in $H$ has been only performed for the small timescales (region (i)), where the values of $H$ are considerably larger and the persistence is more marked.

The spatial distribution of $H$ in both periods of 30 years have been depicted in Fig. 5(c), (d). Both periods show the highest values ($H \in [0.70, 0.79]$) at the northeastern Atlantic coast, which gradually decrease to the southeast and the east. In this last area, the reduction is even more noticeable. The lowest values in both periods correspond to stations No. 20, 28 and 29. The analysis of changes between periods 1960–1989 and 1990–2019 reveal that there is a little reduction of persistence in most stations of the central and northwestern inland areas and in the southern part. These regions present values of $H \in [0.64, 0.73]$ for the first period and $H \in [0.61, 0.70]$ for the second one.

### 3.3. Parameters of the singularity spectra for different regions for periods 1960–1989 and 1990–2019

The non-monotonic function of $q$ found for the scaling exponent in the region (i) in most stations leads to unreliable singular-ity spectra $f(\alpha)$ when the Eq. (5) is applied. They exhibit a non-parabolic shape with abrupt changes for $q < 0$ and large statistical errors. Therefore, the analysis of changes in parameters that describe the multifractal spectra ($\alpha_0$, $w$ and $B$) has been only performed for the region (iii), i.e., for large timescales. An example of these spectra can be found in Fig. 6(a), (b). Similar parabolic shapes have been obtained for every station. In the given example, the multifractal parameters are shown for both subperiods. $\alpha_0$ increases in the last period and, as a result, the spectrum is shifted to the right. Both curves show widths significantly greater than zero, what proves that signals have a multifractal behavior, although it is more evident for the period 1990 – 2019. Two different behaviors can be found for the asymmetry parameter $B$. For the sake of clarity, the shifted curves of spectra with respect to its maximum were depicted together with its respective second order polynomial fit in Fig. 6(c), (d). Additionally, the second order polynomial functions with the same $A$ and $C$ coefficients and null $B$ coefficient are plotted in the same figures. A comparison between both fits is enough to see that this station had approximately symmetric multifractal spectrum (Fig. 6(c)), which changed to a left-skewed spectrum in the last 30 years, with $B = 0.14 \pm 0.04$ (Fig. 6(d)). As explained previously in Section 2.3, this means that this time series becomes more singular, with loss of "fine-structure". The multifractal parameters for all locations are shown with their statistical errors in Table A.2.

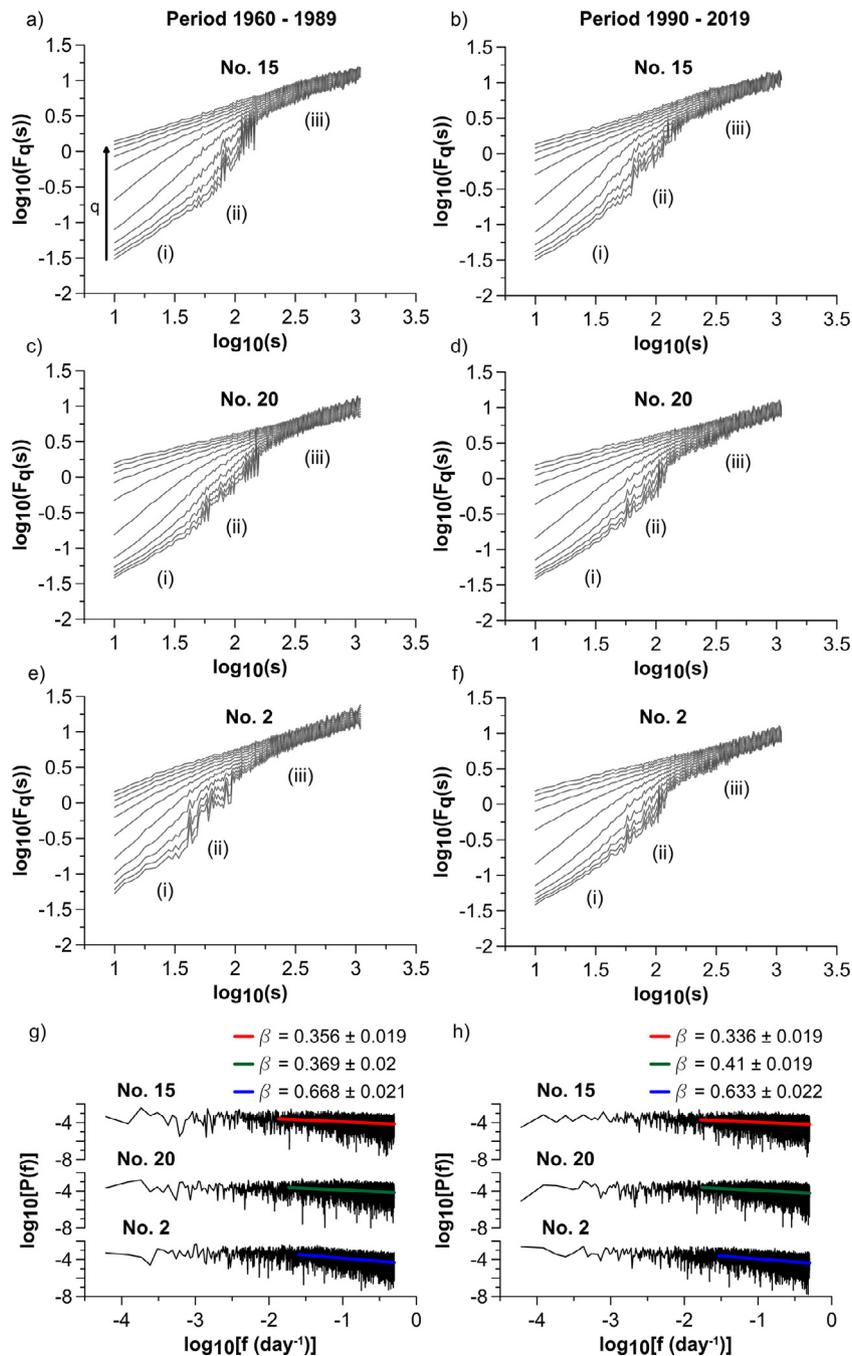

**Fig. 4.** (a–f) Log-log plots of fluctuations function $F_q(s)$ vs time scale $s$ for different stations and periods. For clarity reasons, only half of the analyzed curves are depicted. (g, h) Power spectrum of the same stations and their corresponding slopes.

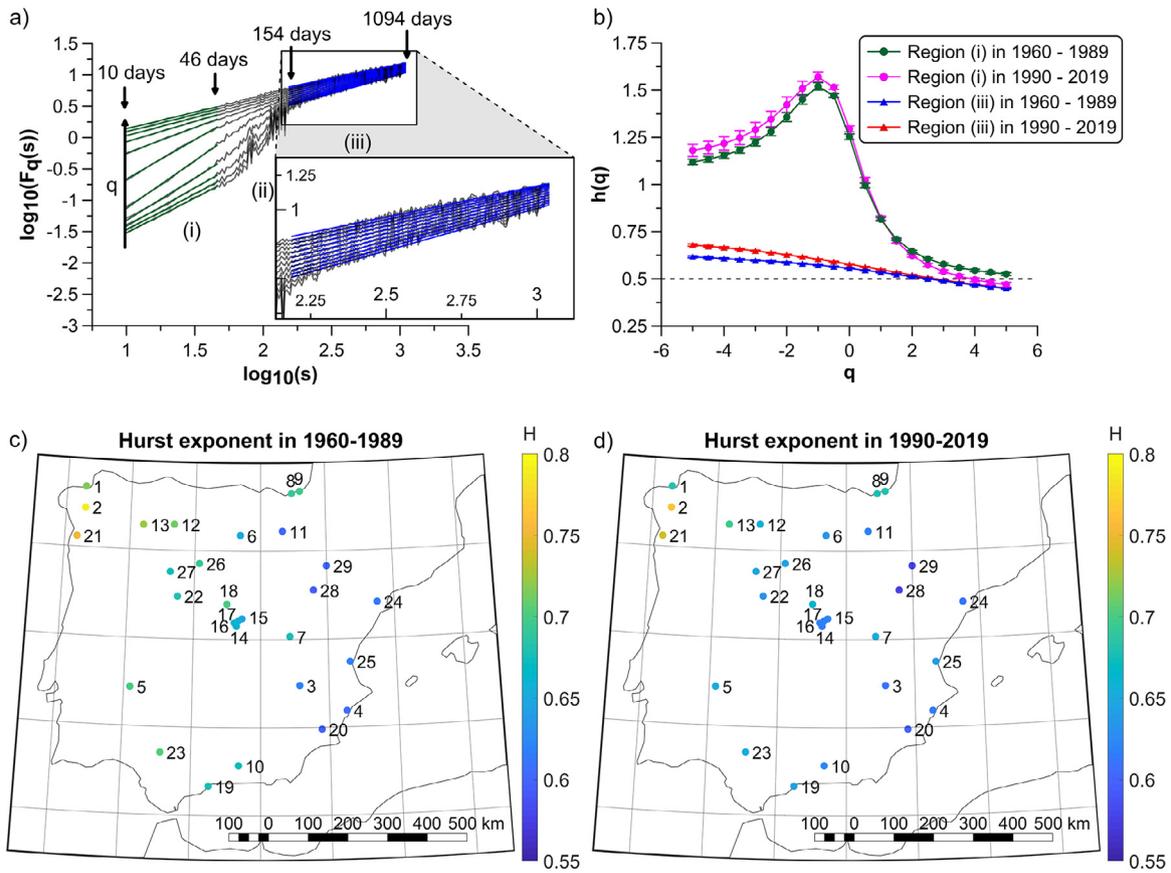

**Fig. 5.** (a) Log-log plot of fluctuation function $F_q(s)$ vs time scale $s$ for different $q$ values in station No. 15 and subperiod 1960 – 1989. (b) $h(q)$ and their respective statistical errors obtained for regions (i) and (iii) for the same station. (c, d) Spatial distribution of $H$ for the region (i).

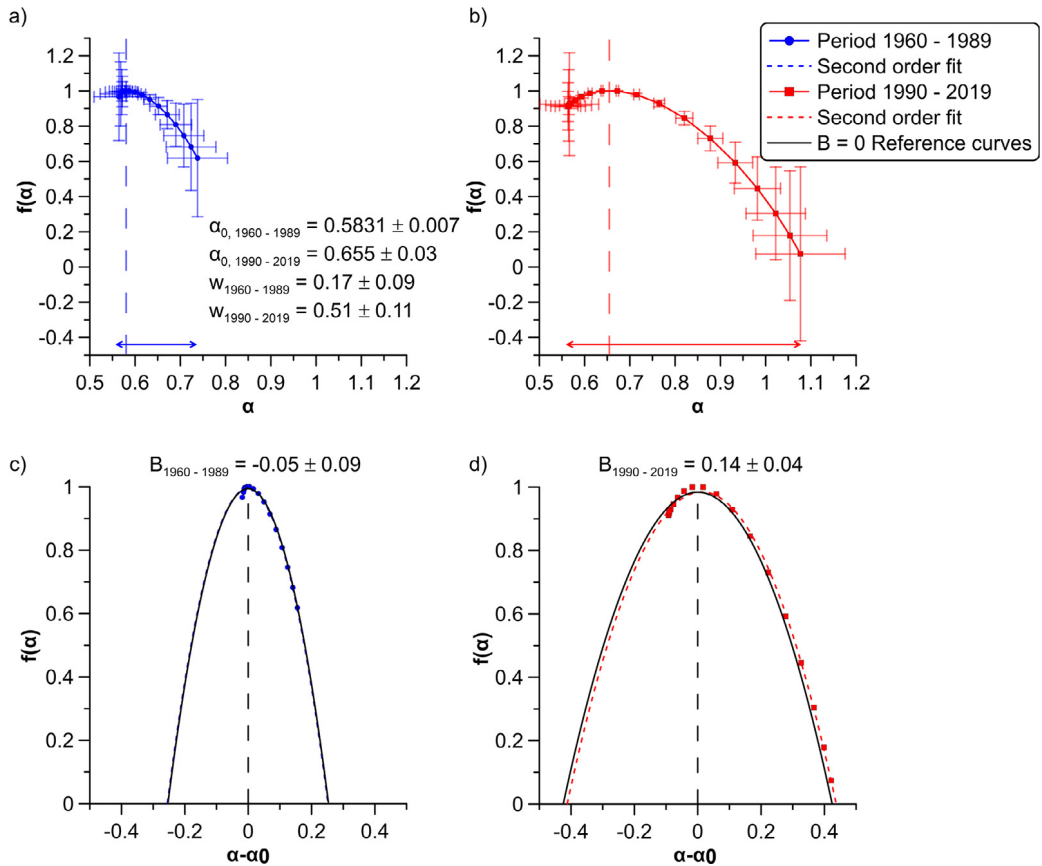

**Fig. 6.** (a, b) Singularity spectra $f(\alpha)$ in station No. 2 in both subperiods obtained for the scaling region (iii). (c, d) Shifted curves of the spectra shown in panels (a, b). The reference curves with the same $A$ and $C$ coefficients from second order fits and $B = 0$ are depicted as solid black lines. For clarity reasons, error bars have been omitted here.

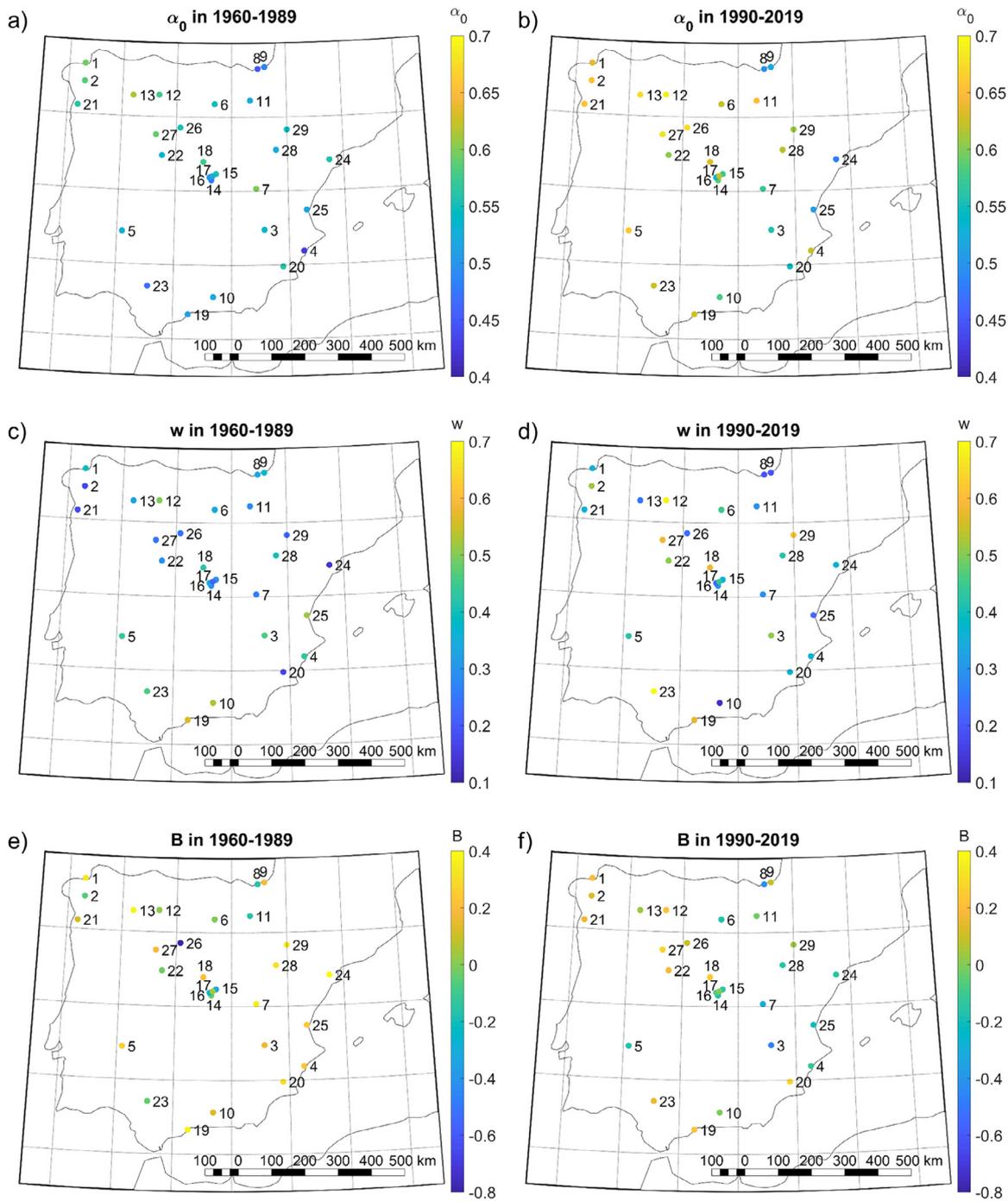

**Fig. 7.** Spatial distribution of multifractal parameters obtained for the scaling region (iii).

The spatial distribution for the maximum scaling exponent $\alpha_0$ for period 1960–1989 is depicted in Fig. 7(a). For this period, there is a high variability in $\alpha_0$. However, some areas which present more homogeneity can be observed. The northwestern coastal part (stations No. 1, 2 and 21), with $\alpha_0 \in [0.56, 0.60]$; the northeastern area of the Iberian Peninsula (stations No. 6, 11, 29 and 28), with $\alpha_0 \in [0.52, 0.55]$; the southern part (No. 19 and 10, $\alpha_0 \sim [0.50, 0.52]$) and the northern Atlantic coast (No. 8 and 9, $\alpha_0 \in [0.45, 0.49]$). This last area shows some of the lowest values of $\alpha_0$, denoting that these series are more characterized by a "smooth-structure". On the contrary, stations No. 13 and 7 are those with the highest value (see Table A.2), meaning that these series are more complex (possess "fine-structure").

On the other hand, the plot for the last 30 years is depicted in Fig. 7(b). There is a slight increase of values for the most part of the sample in the period 1990–2019. The northwestern Atlantic coast present values of $\alpha_0 \in [0.63, 0.66]$, while for the northwestern inland area, $\alpha_0 \in [0.67, 0.69]$; the northeastern area, now excluding station No. 11, which shows values of $\alpha_0 \in [0.60, 0.66]$ and the northern Atlantic coast, with $\alpha_0 \in [0.49, 0.53]$. In this last period, the southern part is more heterogeneous.

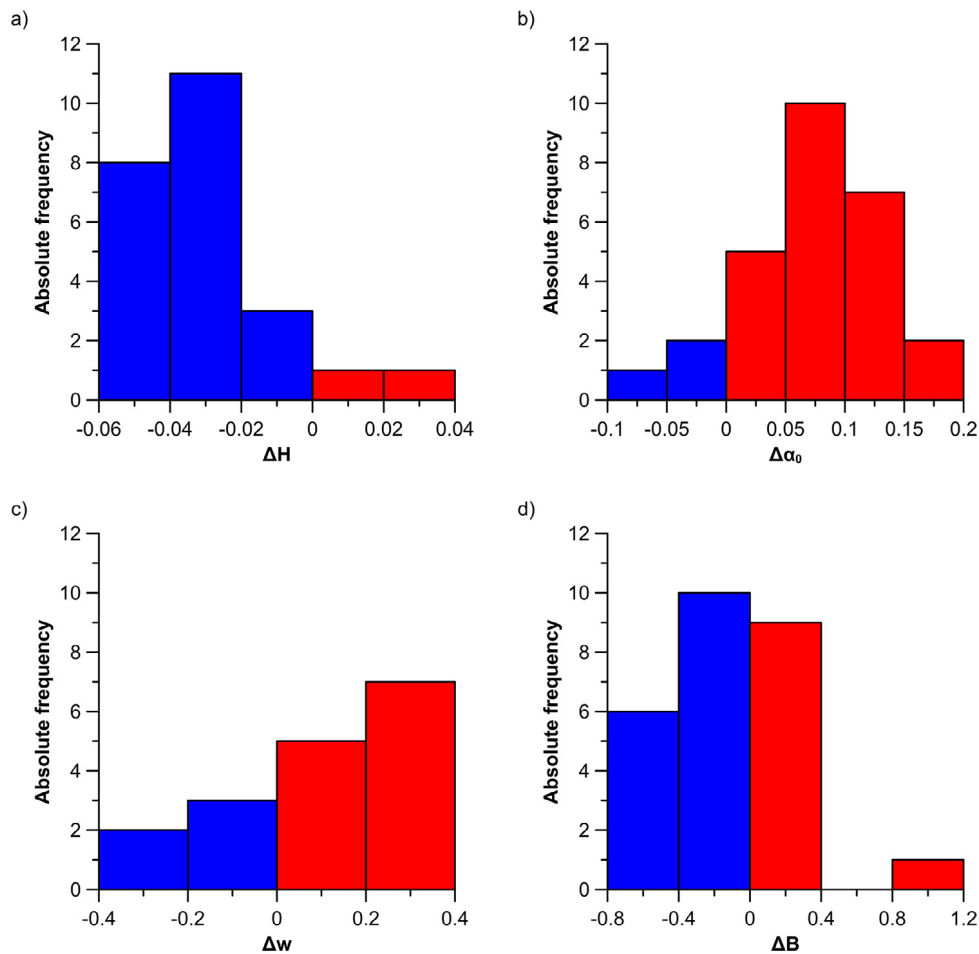

**Fig. 8.** Histograms of changes in parameters between periods 1960 - 1989 and 1990 - 2019. The width of the bins exceeds the largest statistical error in each parameter. Non-significant changes are omitted. (a) Hurst exponent in the scaling region (i). (b, c, d) Parameters of the multifractal spectra in the scaling region (iii).

Fig. 7(c), (d) depict the spatial distribution of the spectral width $w$ for both subperiods. This magnitude shows a wide range of values and high spatial variability. Additionally, a little less than half of stations exhibit non-significant changes. Therefore, it cannot be identified any geographical pattern where these changes are consistent. The highest values correspond to stations No. 19 in the first period and No. 23 in the second one. This means that these time series have relatively more scaling exponents for large and low fluctuations in the respective periods and, thus, they possess a higher degree of multifractality. On the contrary, the lowest values (stations No. 24 in the period 1960–1989 and No. 10 in 1990–2019) are not significantly different from zero (see Table A.2). Hence, these signals are monofractals in the respective periods.

For the asymmetry parameter in the first period (see Fig. 7(e)), it can be observed a region with high homogeneity and positive values ($B \in [0.12, 0.39]$). This region covers the most part of the eastern area. As a result, their respective multifractal spectra are left-skewed. This means that series from the eastern part of the Peninsula in this period are characterized by being singular and having low fluctuations with "smooth-structure". Outcomes for the period 1990 - 2019 are depicted in Fig. 7(f). In this figure, the geographic pattern found in the eastern part of the Iberian Peninsula changes differently depending on the location. However, almost all the locations in this subregion show a decrease of the asymmetry coefficient, what means that they become less singular and exhibit a more complex structure of the low fluctuations.

Furthermore, some of these locations (No. 4, 25, 28 and 24) slightly vary its asymmetry from left to right-skewed, while spectra from stations No. 3 and 7 become noticeably right-skewed. Therefore, they display low fluctuations with "fine-structure" in the last period.

### 3.4. Global changes in $H$, $\alpha_0$, $w$, and $B$ between both periods

To clarify the general changes in $H$ and in multifractal parameters $\alpha_0$, $w$ and $B$, their histograms are shown in Fig. 8. There are 22 stations that exhibit a significant decrease of the Hurst exponent between both periods (Fig. 8(a)). This denotes a global reduction of the persistence in the Iberian Peninsula at small timescales. In Fig. 8(b), an increase of $\alpha_0$ is observed for large scales in 24 stations, what indicates a rise in the complexity of local fluctuations of low magnitude. The width of spectra also shows a rise between both periods, but only for 17 stations (see Fig. 8(c)). For this reason, and because no geographical patterns were found for this parameter, it cannot be discarded that these increases are due to the effect of the local conditions. Lastly, the asymmetry coefficient does not exhibit a marked tendency (see Fig. 8(d)).

## 4. Conclusions

48% of gauge stations present statistically significant decreasing trends in annual precipitation series between 1960 and 2019. These outcomes are also consistent with a decrease in mean and standard deviation of annual precipitation between the subperiods 1960–1989 and 1990–2019. Most stations with decreasing annual rainfall records are distributed over the central part of Spain, which is characterized by a strong continental climate with a relatively low precipitation amount (between 400 and 600 mm/year) [10]. Thus, this region is uniformly becoming more arid in the last period (1990 - 2019).

The MF-DFA method reveals that the fluctuations of the daily precipitation series contain three distinct scaling regions. In the intermediate scales, the curves mainly differ for the fluctuations of low magnitude ($q < 0$). The analysis of the power spectrum only exhibits two different regimes, showing clear differences between the small and large timescales. This indicates that the small fluctuations at intermediate scales belong to a transition region between the mentioned regimes. Similar findings of multi-scale multifractality were also found for precipitation with hourly resolution in the United States [41]. In addition, they have also been observed in other complex systems in different fields of knowledge, such as physics [42], traffic flow [43], astronomy [44] or economics [32].

Generalized Hurst exponents depend on $q$ moments in every case, what is a clear evidence of multifractality. The degree of multifractality is higher for small timescales. In these scales, the series exhibit a non-monotonic shape of $h(q)$ for small fluctuations ($q < 0$). This phenomenon might be a result of the influence of the time series resolution on the analysis at the smallest timescales. This led to unreliable results in the computation of the singularity spectra and the study of these ones was performed for large timescales only.

All stations present a strong long-term persistence for small timescales and weak correlations for large timescales. However, these last correlations are poorly confirmed by the power spectra for a few stations only. The standard Hurst exponent computed for small timescales shows a geographical pattern. It approximately follows the strong gradient of annual precipitation which characterizes the Iberian Peninsula. This gradient exhibits large values to the north and northwest and lower values towards the southeast [10]. The persistence is reduced in most stations between both studied periods, what might be related to the annual precipitation decrease found in the trend analysis.

The features of multifractal spectra for large timescales show more complex patterns. In fact, the spatial analysis of results is not coherent among different parameters. The main conclusions drawn are detailed next. $\alpha_0$ results show that there is a global rise in complexity of low fluctuations (more presence of "fine-structure"), mainly in the Atlantic coast and the northern inland area. On the other hand, half of stations exhibit significant variations of the width, $w$, between both periods, although no pattern can be observed. This might denote that the degree of multifractality might be affected by the local conditions. Finally, $B$ only shows consistent variations in the eastern part of the Iberian Peninsula, where the precipitation series become less singular and have more presence of "fine-structure" in the last period. Nevertheless, the strength of these changes varies depending on the location and is particularly notable in some stations from the central area of the Mediterranean coast and inland. This might be related to the change in the frequency and direction of the Mediterranean cyclones that affects this part of the Iberian Peninsula [45].

It has been shown that several scaling regimes with different multifractal features are present in the precipitation series of the Iberian Peninsula. The nonlinear multifractal features are proved to be more complex and irregularly distributed than those derived from the linear analysis. Furthermore, it has been observed that some of these properties change over time in different scales. This fact might reveal an important role of these parameters in the long-term change of the precipitation.

## Declaration of Competing Interest

The authors declare that they have no known competing financial interests or personal relationships that could have appeared to influence the work reported in this paper.

## CRediT authorship contribution statement

**Javier Gómez-Gómez:** Conceptualization, Methodology, Software, Validation, Visualization, Formal analysis, Data curation, Investigation, Writing – original draft. **Rafael Carmona-Cabezas:** Conceptualization, Software, Investigation, Resources. **Elena Sánchez-López:** Conceptualization, Resources, Supervision. **Eduardo Gutiérrez de Ravé:** Project administration, Funding acquisition, Supervision. **Francisco José Jiménez-Hornero:** Project administration, Funding acquisition, Supervision.

## Acknowledgements

The FLAE approach for the sequence of authors is applied in this work. Authors gratefully acknowledge the financial support of the Andalusian Research Plan Group TEP-957, the Research project UCO-1379178 (ERDF Operational Program Framework Andalusia 2014–2020) and the Research Program of the University of Cordoba (2021), Spain. We also thank the Spanish Meteorological Agency ("Agencia Estatal de Meteorología") for providing data records.

## Appendix A

**Table A.1**
Description of gauge stations of daily precipitation data series recorded at the Iberian Peninsula in the period 1960–2019.

| No. | Location | Latitude (°N) | Longitude (°W) | Altitude (m a.s.l.) |
|---|---|---|---|---|
| 1 | La Coruña | 43.37 | 8.42 | 58 |
| 2 | Santiago de Compostela Airport | 42.89 | 8.41 | 370 |
| 3 | Albacete Air Base | 38.95 | 1.86 | 702 |
| 4 | Alicante | 38.37 | 0.49 | 81 |
| 5 | Badajoz Airport | 38.88 | 6.81 | 185 |
| 6 | Burgos Airport | 42.36 | 3.62 | 891 |
| 7 | Cuenca | 40.07 | 2.13 | 948 |
| 8 | San Sebastián, Igueldo | 43.31 | 2.04 | 251 |
| 9 | Hondarribia, Malkarroa | 43.36 | 1.79 | 4 |
| 10 | Granada Air Base | 37.14 | 3.63 | 687 |
| 11 | Logroño Airport | 42.45 | 2.33 | 353 |
| 12 | León, Virgen del Camino | 42.59 | 5.65 | 912 |
| 13 | Ponferrada | 42.56 | 6.60 | 534 |
| 14 | Getafe | 40.30 | 3.72 | 620 |
| 15 | Madrid Airport | 40.47 | 3.56 | 609 |
| 16 | Madrid, Cuatro Vientos | 40.38 | 3.79 | 690 |
| 17 | Madrid, Retiro | 40.41 | 3.68 | 667 |
| 18 | Navacerrada Pass | 40.79 | 4.01 | 1894 |
| 19 | Málaga Airport | 36.67 | 4.48 | 5 |
| 20 | Alcantarilla Air Base | 37.96 | 1.23 | 75 |
| 21 | Vigo Airport | 42.24 | 8.62 | 261 |
| 22 | Salamanca Airport | 40.96 | 5.50 | 790 |
| 23 | Sevilla Airport | 37.42 | 5.88 | 34 |
| 24 | Tortosa | 40.82 | −0.49 | 50 |
| 25 | Valencia | 39.48 | 0.37 | 11 |
| 26 | Valladolid Airport | 41.71 | 4.86 | 846 |
| 27 | Zamora | 41.52 | 5.74 | 656 |
| 28 | Daroca | 41.11 | 1.41 | 779 |
| 29 | Zaragoza Airport | 41.66 | 1.00 | 249 |

**Table A.2**

Multifractal spectra parameters ($\alpha_0$, $w$ and $B$) in the scaling region (iii) and their respective statistical errors in every location for both subperiods. Results highlighted in bold are not significantly different from zero.

| No. | $\alpha_0$ | | $w$ | | $B$ | |
|---|---|---|---|---|---|---|
| | 1960 - 1989 | 1990 - 2019 | 1960 - 1989 | 1990 - 2019 | 1960 - 1989 | 1990 - 2019 |
| 1 | $0.5940 \pm 0.013$ | $0.6356 \pm 0.013$ | $0.40 \pm 0.08$ | $0.35 \pm 0.11$ | $0.34 \pm 0.04$ | $0.20 \pm 0.09$ |
| 2 | $0.5831 \pm 0.007$ | $0.655 \pm 0.03$ | $0.17 \pm 0.09$ | $0.51 \pm 0.11$ | $\mathbf{-0.05 \pm 0.09}$ | $0.14 \pm 0.04$ |
| 3 | $0.5364 \pm 0.008$ | $0.5593 \pm 0.014$ | $0.46 \pm 0.14$ | $0.51 \pm 0.11$ | $\mathbf{0.17 \pm 0.19}$ | $-0.46 \pm 0.08$ |
| 4 | $0.4276 \pm 0.021$ | $0.623 \pm 0.03$ | $0.43 \pm 0.11$ | $0.37 \pm 0.13$ | $0.21 \pm 0.05$ | $-0.114 \pm 0.015$ |
| 5 | $0.5272 \pm 0.019$ | $0.661 \pm 0.03$ | $0.45 \pm 0.10$ | $0.42 \pm 0.12$ | $0.250 \pm 0.023$ | $-0.180 \pm 0.021$ |
| 6 | $0.5485 \pm 0.013$ | $0.6209 \pm 0.021$ | $0.34 \pm 0.09$ | $0.45 \pm 0.08$ | $-0.0083 \pm 0.0020$ | $-0.179 \pm 0.017$ |
| 7 | $0.6004 \pm 0.010$ | $0.5689 \pm 0.011$ | $0.26 \pm 0.08$ | $0.29 \pm 0.10$ | $0.35 \pm 0.08$ | $-0.30 \pm 0.08$ |
| 8 | $0.4501 \pm 0.017$ | $0.4939 \pm 0.006$ | $0.33 \pm 0.08$ | $0.18 \pm 0.06$ | $-0.150 \pm 0.016$ | $-0.41 \pm 0.06$ |
| 9 | $0.4853 \pm 0.015$ | $0.5218 \pm 0.007$ | $0.38 \pm 0.08$ | $0.19 \pm 0.08$ | $0.182 \pm 0.017$ | $0.083 \pm 0.012$ |
| 10 | $0.518 \pm 0.03$ | $0.5820 \pm 0.005$ | $0.53 \pm 0.10$ | $\mathbf{0.14 \pm 0.15}$ | $0.122 \pm 0.014$ | $\mathbf{0.00 \pm 0.09}$ |
| 11 | $0.5210 \pm 0.013$ | $0.6508 \pm 0.016$ | $0.28 \pm 0.10$ | $0.29 \pm 0.10$ | $-0.154 \pm 0.014$ | $-0.028 \pm 0.013$ |
| 12 | $0.5739 \pm 0.019$ | $0.689 \pm 0.03$ | $0.50 \pm 0.09$ | $0.68 \pm 0.10$ | $0.010 \pm 0.004$ | $0.217 \pm 0.018$ |
| 13 | $0.6145 \pm 0.014$ | $0.6723 \pm 0.013$ | $0.33 \pm 0.09$ | $0.24 \pm 0.06$ | $0.37 \pm 0.05$ | $0.038 \pm 0.020$ |
| 14 | $0.4769 \pm 0.014$ | $0.5880 \pm 0.010$ | $0.29 \pm 0.08$ | $0.30 \pm 0.11$ | $-0.049 \pm 0.009$ | $-0.17 \pm 0.10$ |
| 15 | $0.5549 \pm 0.014$ | $0.5771 \pm 0.020$ | $0.29 \pm 0.08$ | $0.37 \pm 0.10$ | $-0.33 \pm 0.04$ | $-0.128 \pm 0.015$ |
| 16 | $0.4989 \pm 0.017$ | $0.5336 \pm 0.010$ | $0.35 \pm 0.07$ | $0.22 \pm 0.11$ | $-0.26 \pm 0.03$ | $-0.082 \pm 0.009$ |
| 17 | $0.5195 \pm 0.008$ | $0.6206 \pm 0.021$ | $0.22 \pm 0.09$ | $0.44 \pm 0.11$ | $\mathbf{0.04 \pm 0.07}$ | $\mathbf{-0.003 \pm 0.008}$ |
| 18 | $0.5759 \pm 0.021$ | $0.630 \pm 0.03$ | $0.42 \pm 0.16$ | $0.58 \pm 0.09$ | $0.199 \pm 0.020$ | $0.25 \pm 0.04$ |
| 19 | $0.5098 \pm 0.024$ | $0.627 \pm 0.05$ | $0.57 \pm 0.11$ | $0.58 \pm 0.16$ | $0.37 \pm 0.06$ | $0.21 \pm 0.03$ |
| 20 | $0.5610 \pm 0.004$ | $0.5410 \pm 0.016$ | $0.17 \pm 0.12$ | $0.37 \pm 0.10$ | $0.29 \pm 0.15$ | $0.27 \pm 0.06$ |
| 21 | $0.5640 \pm 0.006$ | $0.6562 \pm 0.018$ | $0.18 \pm 0.09$ | $0.36 \pm 0.12$ | $0.11 \pm 0.10$ | $0.187 \pm 0.019$ |
| 22 | $0.5317 \pm 0.013$ | $0.6056 \pm 0.022$ | $0.29 \pm 0.08$ | $0.51 \pm 0.09$ | $-0.030 \pm 0.007$ | $0.132 \pm 0.019$ |
| 23 | $0.4658 \pm 0.019$ | $0.621 \pm 0.04$ | $0.46 \pm 0.15$ | $0.73 \pm 0.11$ | $-0.044 \pm 0.011$ | $0.159 \pm 0.017$ |
| 24 | $0.5573 \pm 0.017$ | $0.4808 \pm 0.016$ | $\mathbf{0.14 \pm 0.17}$ | $0.36 \pm 0.12$ | $0.4 \pm 0.4$ | $-0.132 \pm 0.011$ |
| 25 | $0.5146 \pm 0.022$ | $0.5083 \pm 0.012$ | $0.52 \pm 0.13$ | $0.23 \pm 0.09$ | $0.24 \pm 0.05$ | $-0.21 \pm 0.04$ |
| 26 | $0.5504 \pm 0.013$ | $0.6720 \pm 0.009$ | $0.24 \pm 0.12$ | $0.23 \pm 0.12$ | $-0.8 \pm 0.3$ | $0.089 \pm 0.017$ |
| 27 | $0.5904 \pm 0.009$ | $0.6792 \pm 0.024$ | $0.25 \pm 0.09$ | $0.58 \pm 0.11$ | $0.19 \pm 0.04$ | $0.27 \pm 0.06$ |
| 28 | $0.5254 \pm 0.015$ | $0.6194 \pm 0.016$ | $0.41 \pm 0.11$ | $0.42 \pm 0.08$ | $0.32 \pm 0.06$ | $-0.146 \pm 0.016$ |
| 29 | $0.5488 \pm 0.010$ | $0.6071 \pm 0.020$ | $0.23 \pm 0.09$ | $0.61 \pm 0.09$ | $0.34 \pm 0.06$ | $0.037 \pm 0.008$ |